\documentclass{moriond}

\def\be{\begin{equation}}
\def\ee{\end{equation}}
\def\bea{\begin{eqnarray}}
\def\eea{\end{eqnarray}}

\newcommand{\Photo}{}

\begin{document}
\vspace*{4cm}
\title{Gravitational wave alert follow-up strategy in the H.E.S.S. multi-messenger framework}

\author{ M. Seglar-Arroyo and F. Sch\" ussler on behalf of the H.E.S.S. Collaboration}

\address{IRFU, CEA,
Universit\'e Paris-Saclay,\\
F-91191 Gif-sur-Yvette France\\	Contact e-mail: monica.seglar-arroyo@cea.fr }

\maketitle\abstracts{The H.E.S.S. high-energy gamma-ray observatory is member of the Virgo/LIGO electromagnetic follow-up effort since early 2014. Its capability for transient follow-up studies benefits from its large field of view, rapid response time and high sensitivity. Drawing from the experience gained from other science cases like gamma-ray bursts and high-energy neutrino follow-ups we demonstrate the high perspectives for new types of analyses like the search for gravitational wave counterparts and the study of multi-messenger signals from binary neutron star mergers. This contribution aims to present the potential pointing strategy that the H.E.S.S. observatory would carry out following an alert from gravitational wave observatories. We will discuss several key points like the use of information from a galaxy catalogue, the time-dependent visibility of sky regions and the automatic handling of gravitational wave uncertainty maps, that will enable an optimized choice of the pointing directions. Finally, based on simulated binary neutron star mergers, the performance of the outlined gravitational wave-alert observations will be presented. }

\section{Introduction}\label{intro}

The continuous study of the high-energy sky has revealed a large number of powerful astrophysical objects capable of emitting radiation in the entire electromagnetic (EM) spectrum. Recent discoveries of high-energy astrophysical neutrinos and the direct detection of gravitational waves (GW) not only have opened new ways to explore the sky but also started a new era in the understanding of the most energetic phenomena in the Universe by means of combining different cosmic messengers. 

This multi-messenger astronomy aims to solve longstanding puzzles, as the origin of cosmic rays, by linking different observations from instruments sensitive to various astrophysical messengers as neutrinos, cosmic rays, electromagnetic radiation and gravitational waves. Indeed, the study of the non-electromagnetic messengers is currently done in facilities with sensitivities naturally limited,  in the case of neutrinos due to its low cross section and, for gravitational waves, due to the extreme weakness of the signal. The goal in setting up follow-up strategies is to beneficiate of the electromagnetic radiation observatories experience and event rates.


\subsection{Gravitational waves}
\label{Gravitational waves}

Einstein's theory of general relativity predicts the propagation of fluctuations in the metric of spacetime as gravitational waves\cite{einstein}. The existence of gravitational waves was indirectly proven by over three decades of measurements of the orbit of the binary pulsar PSR1912+16 \cite{GWindirectly}, but this fundamental prediction of general relativity had not been tested directly. Even though their direct detection stands as a major challenge, current technology of detection is sensitive enough to make the first direct measurement.


Highly compact astrophysical objects with non-symmetric mass distribution are expected to be source of gravitational waves by distorting spacetime. If this effect is violent enough and if the wavelength of the propagating wave is in the correct range, it may be detected by current instruments. The most promising astrophysical GW sources in the frequency band of current detectors aLigo/aVirgo are the inspiral and coalescence of compact binaries with neutron stars and/or black holes (BH) constituents \cite{Berger}. When the orbit of the binary steadily decays as a result of the gravitational wave emission, it causes the astrophysical objects to spiral together at a constantly increasing rate as the merger approaches. In many cases, their gravitational waveform is expected to contain many cycles in the sensitive band of the detectors. In other cases with different mass, it is the merge who is expected to create, a wave that could be detected from Earth. The expected number of detections that could be made for such mergers varies, however, by up to four orders of magnitude \cite{prospects}.

The study of the transient sky of short-duration cataclysmic events leading to GW signals has just started with phenomena as stellar core-collapse, gamma-ray burst engines, and mergers of compact object binaries. However, in order to maximise the scientific potential of such discoveries, complementary EM data are needed.

Joint GW and EM broadband observation would allow to combine measurements of masses, distance and binary inclination among other parameters given by the GW reconstruction \cite{SingerReconstruction} with luminosity, redshifts and duration measurements available from EM observations. This will enable to constraint accurately the energies, physics of the merger event and locations in the sky that may allow the identification of the host galaxy and source environment, allowing to tell a quite complete story of the phenomenon that would be unaccesible otherwise.

\subsection{GW observatories and improvements on reconstruction}
\label{subsec:GW observatories and improvements on reconstruction}

The first detection of the first gravitational wave signal from a binary black hole merger was done by the collaboration LIGO-Virgo on September 2015 \cite{GWfirstdetection}, during its first science run (O1), almost inmediately after start operating with the upgraded interferometer, known as Advanced LIGO (aLIGO). Since then, searching for electromagnetic counterparts to these events became a great challenge and big efforts have been done in order to look for transient signatures. A partnership between multi-wavelenght facilities from gamma-ray to radio have joined the effort. This campaign allowed to perform extensive EM follow-up searches of the first GW event, GW150914 \cite{GW150914follow-up}.

However, the GW detector network performance depends on the number of detectors, their geographical distribution, the relative orientation of the detector arms with respect to the incoming GW and their relative sensitivity. With the current setup of two LIGO interferometers, the 90\% credible regions could be as large as thousands of squared degrees \cite{prospects} due to the lack of triangulation and hence, the identification of an electromagnetic counterpart becomes non trivial. Reducing the sky position of a significant fraction of detected signal to areas of several tens of degrees requires a third detector of sensitivity within a factor of $\sim$  2 for each other and a broad frequency bandwidth. 

The improvement of response to GW alerts depends on both, reconstruction of the GW event and the follow-up strategy set up by the follow-up observatory.

Concerning the GW side, improvements in reconstruction have been done in order to include 3D information in the O2 run period that started November 2016. A rapid GW parameter estimation algorithm reconstructing the full distribution of sky locations and distance was a key missing ingredient for EM-counterpart follow-up observations.\cite{goingthedistance} 

Regarding the EM-experiment side, the area on the sky that must be searched for an EM-counterpart may be reduced by combining the reconstructed volume with position and redshift of galaxies that could host such a signal, and eventually other information as mass, luminosity and star formation rate that is available from galaxy catalogues. This permits the optimisation and improvement of follow-up searches by correlating with a galaxy catalogue as it has been shown in previous studies \cite{swift2015}.

\subsection{Follow-up programs in H.E.S.S.}
\label{subsec:Follow-up programs in H.E.S.S.}

A well-suited instrument for transient searches should present several key features that enables an efficient response to transient events. Among them, the instruments should present rapid follow-up response, high sensitivity and large Field-of-View (FoV) for exploring larger areas in the sky. In the very high-energy gamma-ray domain, H.E.S.S. experiment stands out as a powerful instrument capable of performant follow-up searches.

H.E.S.S. is a stereoscopic system consisting of five imaging atmospheric Cherenkov telescopes located at 1800 meters above sea level in the Khomas highlands of Namibia (23$^{\circ}$16’ 18” South, 16$^{\circ}$ 30’ 00” East). The four H.E.S.S. phase I telescopes are arranged in a square of 120 m side length, with the large H.E.S.S. II telescope at its centre. In its initial four-telecope configuration, the sensitivity to gamma-ray energies from 100 GeV to about 100 TeV while the energy threshold of the H.E.S.S. II phase is $\sim$ 30 GeV and angular resolution smaller than 0.1$^{\circ}$-0.4$^{\circ}$.

The H.E.S.S. experiment already has implemented dedicated Target-of-Opportunity programs in Gamma-Ray Burst (GRB)\cite{Tania}, Fast Radio Burst (FRB) and multi-wavelenght AGN studies \cite{Jill}. Follow-up searches with other cosmic messengers as high-energy neutrinos confirm its performance and capabilities. For details on the multi-messenger program see \cite{Fabian}. In the field of gravitational waves, the H.E.S.S. collaboration is member of the Virgo/LIGO EM-follow-up effort since early 2014.
 
\section{Observation scheduler}
\label{Observation scheduler}

To perform follow-up observations of GW events we developed a dedicated scheduling tool. This scheduler is using the following ingredients: 

\begin{itemize}
\item Gravitational wave map of the signal that has been detected, reconstructed and distributed by the GW experiment, in this case aLIGO \cite{ligo}.

\begin{figure}[h]
\centering
\includegraphics[width=\textwidth,clip]{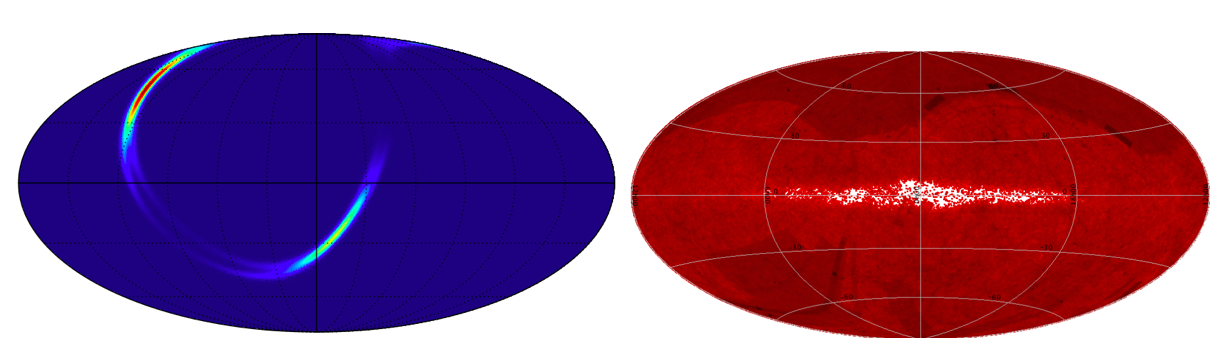}
\caption{Gravitational wave signal of a simulated binary neutron star (NS) merger provided by the LIGO-Virgo collaboration (left).  \texttt{GLADE} galaxy catalogue (right)}
\label{fig-2}       
\end{figure}

\item Galaxy catalog. The Galaxy List for the Advanced Detector Era\cite{glade}  (\texttt{GLADE}) catalog is a value-added full-sky galaxy catalog highly complete and built in order to support EM follow-up when searching for GW signal sources. It has been combined and matched from other galaxy catalogues, namely GWGC, 2MPZ, 2MASS XSC and HyperLEDA and also the SDSS-DR12 quasar catalog. It includes more than 3 million entries and it is complete up to $\sim$ 70 Mpc in terms of blue luminosity and about $\sim$50 \% up to 300 Mpc. Observations suggest that the probability of a compact binary coalescing is influenced by recent star formation, what encourages the use of the B-band luminosity completeness as an indicator of higher probability of the galaxy hosting a GRB event. 

\item H.E.S.S. experiment conditions. Coordinates of the observatory, dark periods and visibility conditions ($\theta_{zenith}<45^{\circ}$) are taken into account. The choice is motivated from the fact that lowering the zenith angle $\theta_{zenith}$ translates in lowering the energy threshold of the observation. 
\end{itemize}

\section{Pointing algorithms}
\label{Pointing algorithms}

The performance of the pointing algorithm can be expressed in terms of different types of probability, depending on the observable of interest and related to the source of the gravitational signal one is aiming to study. 

\begin{itemize}
\item Probability of the GW signal - P$_{GW}$. The GW probability map sent by the LIGO-Virgo collaboration with contours related to different uncertainty regions of the signal, from 10\% to 90\% probability regions, as seen in Figure \ref{fig-8}. This allows a 2D treatment of the probability.
\item Probability weighted by potential sources - P$_{GG}$. It uses the distance information included in the new reconstruction algorithm provided in the gravitational wave map distributed by the GW collaboration as presented in \ref{subsec:GW observatories and improvements on reconstruction}. The three projections of the probability distribution can be seen in Figure \ref{fig-8}. Then, this is combined with galaxy catalog information, concretely by taking the distance of the object, what gives a 3D posterior probability distribution that correlates information about the GW event and galaxies as described in \cite{goingthedistance}.   

\begin{figure}[h]
\centering
\includegraphics[width=7.6cm, height=5.9cm,clip]{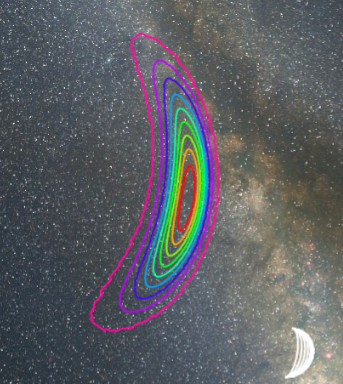}
\includegraphics[width=7.8cm, height=6.1cm,clip]{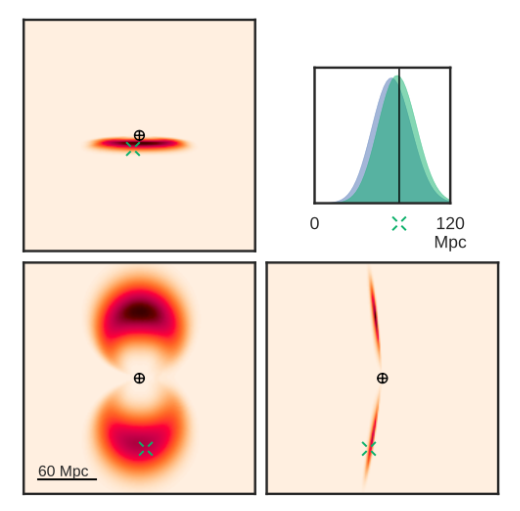}
\caption{Contours containing different probability density regions from 90\% to 10\% of the GW probability map given by the Ligo-VIRGO collaboration (left). Marginal posterior probability distribution in the principal planes from $^6$ (right). The inset in the upper right panel shows the marginal distance posterior distibution integrated over the whole sky (blue) and in the true direction of the source (green).}
\label{fig-8}       
\end{figure}
\end{itemize}

Different algorithms have been developped that consider both presented probabilities in different ways. The approach is related to the science case and thus, the pointing directions prefered by the scheduler.

\subsection{One-galaxy approach}
\label{One-galaxy approach}
The strategy is based on pointing observations according to select individual high probability galaxies and observe them one-by-one. This is done iteratively and in each iteration the objects present in the FoV area are substracted before re-calculating the following pointing position. The goal that presents this first approach is the speed of computation that is a key point for transients searches. However, the coordinates of the pointing with the highest probability value are eventually right next to the FoV of the last pointing, as can be seen in Figure \ref{fig-7}. Due to these overlapping regions the achieved final coverage of the GW error map is, thus, not optimal. For this reason, another approach has been developped and it is introduced in the following subsection.\\

\subsection{Galaxies-in-FoV approach}
\label{Galaxies-in-FoV approach}

\begin{figure}[h]
\centering
\includegraphics[width=\textwidth,height=7.4cm,clip]{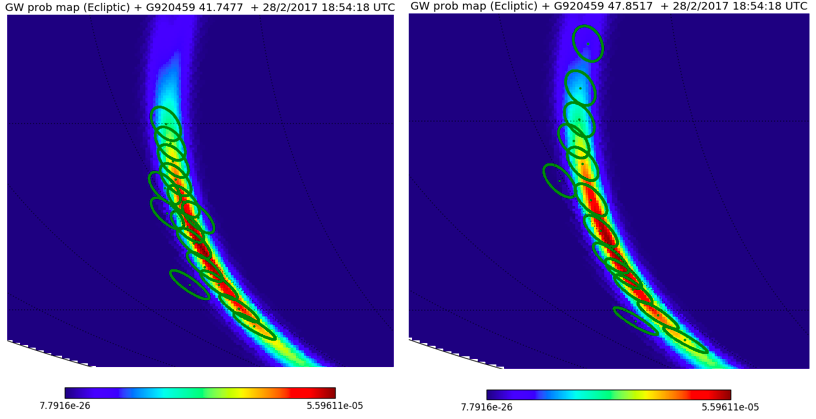}
\caption{Pointing strategy for a gravitational wave signal of simulated NS-NS merger for a  random chosen time with the one-galaxy approach (left) and the galaxies-in-FoV approach (right)}
\label{fig-7}       
\end{figure}

The strategy is based on the optimisation of the pointing strategy in term of the P$_{GG}$ quantity. In addition, instead of individual galaxies, we here take into account the full FoV. The total probability of individual regions defined by the FoV is computed, sorted and the one presenting the highest value is chosen to be observed. The algorithm is also iterative and in every step the galaxies already observed are substracted. The goal of this approach is that it is clearly more performant, since its choses regions with maximal P$_{GG}$  and avoids useless overlapping. The criterium is to maximise the P$_{GG}$ quantity of each pointing, so in some cases some overlapping can be seen in the right plot of Figure \ref{fig-7} since even when some part inside the FoV is not taken into account in the sorting of most probable regions, it still maximises P$_{GG}$. Hence, this approach improves both types of presented probability that yields an increase of about 10\% for classical banana-shape signals, but still, clearly dependent on the shape of the uncertainty region of the GW map that is received. 

However, the probability calculation of the region is currently too time-consuming  for real-time follow-up scheduling due to the size of the catalogue. Various ways to improve the calculation speed are under investigation.

\section{Coverage simulations}
\label{Coverage simulations}
In order to simulate the H.E.S.S. follow-up of GW events and explore the performance of the presented algorithms, coverage simulations have been carried out. Simulations are done for the H.E.S.S. experimental configuration, random arrival times during the year and the following inputs: 

\begin{itemize}
\item Effective FoV covered by a single pointing is a circle of radius of 2.5 degrees, since the acceptance of the pointing is radial. Corresponding to the H.E.S.S. preliminary strategy for GW follow-up studies, a maximum of 20 observation windows of 30 mins (within 3 days) is allowed. 
\item 250 available gravitational wave localisation maps derived from simulated NS-NS merger events provided by the LIGO-Virgo collaboration\cite{ligo}. Sky maps for compact binary merger candidates are produced by two different codes. Here, the online reconstruction algorithm, known as the BayeStar algorithm\cite{goingthedistance}, has been chosen. Note that those simulations were thought to be representative for the O2 run but indeed they have to be considered optimistic due to the delay in the Virgo commissioning.
\item \texttt{GLADE} galaxy catalog presented in Sec. \ref{Observation scheduler}
\end{itemize}

Simulations are done pursuant the following steps:

\begin{enumerate}
\item  Inject the simulated GW maps 10 times at random dates.
\item Calculate the full pointing strategy for both algorithms taking into account visibility and darktime conditions.
\item Derive the achieved coverage for each simulated follow-up.
\item Derive the average coverage.
\end{enumerate}

\begin{figure}[h]
\centering
\includegraphics[width=7.9cm,clip]{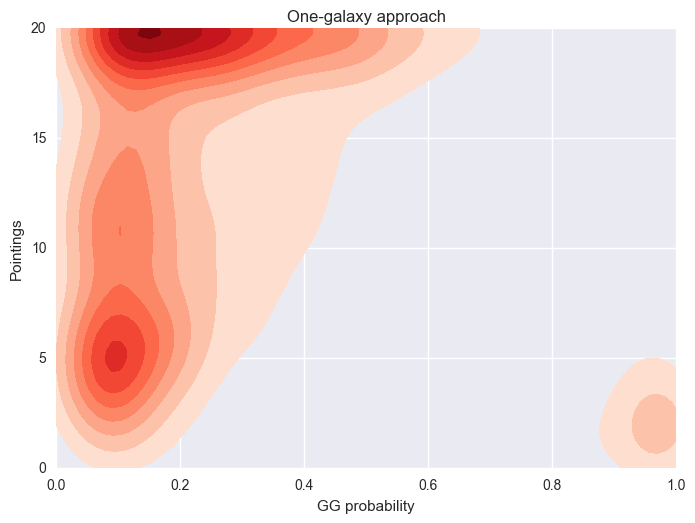}
\includegraphics[width=7.9cm,clip]{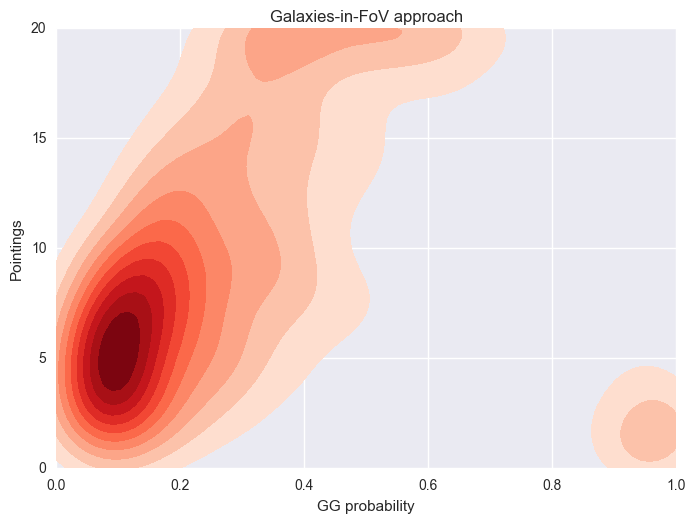}
\caption{Coverage simuations for P$_{GG}$ using the one-galaxy approach (left) and the galaxies-in-FoV approach  (right)}
\label{fig-4}      
\end{figure}

As a result of this simulation process, changes of about 10\% are observed in mean and median covered probability for both of the presented probabilities when chosing Galaxies-in-FoV approach. Though, the big change appears in the number of pointings needed for covering a certain amount of the P$_{GG}$ probability \footnote{Note that the P$_{GW}$ probability presents similar behaviour as seen in Figure \ref{fig-4} when changing the approach}. Figure \ref{fig-4} shows the 2D histogram of the P$_{GG}$ as a function of the number of pointings for both approaches. The distributions in the Galaxies-in-FoV approach case is concentrated in small number of pointings. In general terms, the increase of the total covered probability with the number of pointings is larger for this approach, being related to the overlapping effect seen in Figure \ref{fig-7}. 

Note that the simulated NS-NS merger maps that have been used present large uncertainty in the localisations that expands to both, northern and southern sky, preventing a higher probability coverage as some parts of the sky are not reachable for the experiment. This translates into a limit in Figure \ref{fig-4} while going to large number of pointings.
 
However, one realises about the existance of the spot in the low number of pointings but high coverage part of both plots. This effect is due to the several NS-NS input maps that present a really accurate localization in a convenient part of the sky for the H.E.S.S. experiment, so that they will be almost fully covered in a few pointings.

\section{Summary and Outlook}
\label{summary}
The era of gravitational wave astronomy has just started and thus, a new messenger has been added to multimessenger astronomy. In this contribution we presented the new GW follow-up strategy of the H.E.S.S collaboration. We discussed the use of galaxy catalogues to prioritise sky areas and different definitions of the covered GW localisation uncertainty. We concentrated in developping and comparing different algorithms and open the field for further improvements.

By combining a galaxy catalog with the distributed GW uncertainty region map and accounting for the observatory conditions, an observation scheduler has been set up. As an output, it returns the total probability coverage in terms of GW signal but also of the probability related to galaxy distribution, probability covered in every pointing, observation times and coordinates of the pointing. The algorithms vary depending on the science case and the strategy:

\begin{itemize}
\item One-galaxy approach.  Since it takes only $\sim$ minutes to run it, it is convenient for rapid response during the night. It has, therefore, been implemented in the fully automatic alert reaction scheme running at the site of the H.E.S.S. observatory in Namibia. For details see \cite{Fabian}. 	
\item Galaxies-in-FoV approach. Due to its good performance and its optimisation in terms of probability, this strategy is the most interesting since less pointings are needed to achieve good probability coverage. However, giving a detailed schedule up to 3 days takes currently too much time for automatic reaction. We forsee to employ this algorithm for offline scheduling, i.e. when the GW alerts arrives outside the operation window of H.E.S.S. and for refining the online algorithm. 
\end{itemize}

When using a galaxy catalogue that presents a large number of entries as the \texttt{GLADE} catalogue, we risk to increase too much the calculation time of the algorithm. Therefore strategies to combine both approaches depending on the science case and the type of GW event are being considered. In this context we will for example investigate further use of the information provided by the catalogue (e.g. the blue luminosity of the galaxies with an aim to prefer regions of high star formation rate, etc.)

\begin{figure}[h]
\centering
\includegraphics[width=\textwidth,clip]{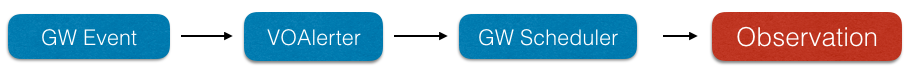}
\caption{Chain of the gravitational wave strategy at the H.E.S.S. experiment}
\label{fig-5}      
\end{figure}

This work is part of the H.E.S.S. transient strategy of using a fully automated chain at the site of the experiment, presented in Figure \ref{fig-5}, developped in order to improve the performance and time reponse to transient events. 

Efforts presented in this work are just a probe of the performance of the electromagnetic follow-up searches that have just started. Upcoming experiments like CTA will take advantage of the expertise gained in current observatories and technology in development that will presumably yield to great achievement and new discoveries. Together with it, note that multi-messenger astronomy will beneficiate in several years of the reduction of the location uncertainty of the GW signal. Currently, GW localizations are $\sim$ 100-1000 deg$^2$ and should shrink to $\sim$ 10-100 deg$^2$  in the future with the detector upgrades and new detectors as KAGRA and LIGO-India \cite{prospects}.

\section*{Acknowledgments}

The author gratefully acknowledge the Moriond Organization Comitee for the financial support that made this contribution possible. The support of the Namibian authorities and of the University of Namibia in facilitating the construction and operation of H.E.S.S. is gratefully acknowledged, as is the support by the German Ministry for Education and Research (BMBF), the Max Planck Society, the German Research Foundation (DFG), the French Ministry for Research, the CNRS-IN2P3 and the Astroparticle Interdisciplinary Programme of the CNRS, the U.K. Science and Technology Facilities Council (STFC), the IPNP of the Charles University, the Czech Science Foundation, the Polish Ministry of Science and Higher Education, the South African Department of Science and Technology and National Research Foundation, and by the University of Namibia. We appreciate the excellent work of the technical support staff in Berlin, Durham, Hamburg, Heidelberg, Palaiseau, Paris, Saclay, and in Namibia in the construction and operation of the equipment.



\end{document}